\documentclass[%
 reprint,
nofootinbib,
 amsmath,amssymb,
 aps,
]{revtex4-2}

\usepackage{graphicx}
\usepackage{dcolumn}
\usepackage{bm}
\usepackage{hyperref}

\def\nm{\nonumber}

\def\pls{{\rm (p)}}
\def\mns{{\rm (n)}}

\def\ampp{\hat{\cal A}^{\pls}}
\def\ampn{\hat{\cal A}^{\mns}}

\def\vecx{{\boldsymbol x}}
\def\veck{{\boldsymbol k}}

\def\hn{{\boldsymbol n}}
\def\hk{{\boldsymbol n}_{\veck}}

\usepackage{color}

\begin{document}

\title{Antipodal Angular Correlations of \\
Inflationary Stochastic Gravitational Wave Background}

\author{Zhen-Yuan Wu}
\affiliation{Graduate School of Sciences and Technology for Innovation, Yamaguchi University, Yamaguchi 753-8512, Japan}

\author{Ryo Saito}
\affiliation{Graduate School of Sciences and Technology for Innovation, Yamaguchi University, Yamaguchi 753-8512, Japan}
\affiliation{Kavli Institute for the Physics and Mathematics of the Universe (WPI), University of Tokyo Institutes for Advanced Study, University of Tokyo, Chiba 277-8583, Japan}

\author{Nobuyuki Sakai}
\affiliation{Graduate School of Sciences and Technology for Innovation, Yamaguchi University, Yamaguchi 753-8512, Japan}

\date{\today}

\begin{abstract}
The measurement of the inflationary stochastic gravitational-wave background (SGWB) is one of the main goals of future GW experiments. 
In direct GW experiments, 
an obstacle to achieving it is the isolation of the inflationary SGWB from the other types of SGWB. 
In this paper, as a distinguishable signature of the inflationary SGWB, we argue the detectability of its universal property: antipodal correlations, i.e., correlations of GWs from the opposite directions, as a consequence of the horizon re-entry. 
A phase-coherent method has been known to be of no use for detecting the angular correlations in SGWB due to a problematic phase factor that erases the signal. 
We thus investigate whether we can construct a phase-incoherent estimator of the antipodal correlations in the intensity map. 
We found that the conclusion depends on whether the inflationary GWs have statistical isotropy or not. 
In the standard inflationary models with statistical homogeneity and isotropy, there is no estimator that is sensitive to the antipodal correlations but does not suffer from the problematic phase factor. 
On the other hand, it is possible to find a non-vanishing estimator of the antipodal correlations for inflationary models with statistical anisotropy. 
SGWB from {\it anisotropic} inflation is distinguishable from the other components.
\end{abstract}

\maketitle

\section{\label{sec:introduction} Introduction}

Inflation \cite{Starobinsky:1980te,Kazanas:1980tx,Sato:1980yn,Guth:1980zm} is a strong candidate for the mechanism to seed the structure of our universe. 
According to the standard paradigm, 
the accelerated expansion during inflation stretches the microscopic quantum fluctuations of the inflaton field to superhorizon scales, which are converted to the primordial density fluctuations in the post-inflationary universe. 
An inevitable prediction is that inflation also generates the primordial gravitational waves (GWs) from tensor-type quantum fluctuations of spacetime \cite{Grishchuk:1974ny, Starobinsky:1979ty, Rubakov:1982df}. 
Thus, the detection of the inflationary GWs gives strong evidence for inflation. 

In direct-detection experiments, 
the inflationary GWs are observed as a stochastic gravitational-wave background (SGWB), i.e., GWs coming from all directions in the sky. 
The detection of the inflationary SGWB is challenging because it has a tiny amplitude in typical inflationary models. 
As well as improving the sensitivity of GW detectors, 
we need to isolate it from SGWB generated by the other sources: a superposition of GWs from many unresolvable astrophysical and cosmological sources (see, e.g., Refs.~\cite{Regimbau:2011rp, LIGOScientific:2016jlg, amaro2017laser, Christensen:2018iqi, Caprini:2018mtu,  LIGOScientific:2019vic}). 
Numerous studies have been carried out on methods for separating the astrophysical components in SGWB: spectral separation \cite{Ungarelli:2003ty,Adams:2013qma, Parida:2015fma,Pieroni:2020rob,Boileau:2020rpg,Poletti:2021ytu}, subtraction \cite{Regimbau:2016ike, Pan:2019uyn, Sharma:2020btq, Martinovic:2020hru, Sachdev:2020bkk}, anisotropies
\cite{Cusin:2017fwz, Adshead:2020bji,Malhotra:2020ket,LISA:2022kbp,Dimastrogiovanni:2019bfl,Dimastrogiovanni:2021mfs,Dimastrogiovanni:2022afr,Orlando:2022rih}, 
polarization \cite{Seto:2006dz,Seto:2006hf,Smith:2016jqs, Domcke:2019zls, Seto:2020zxw, Orlando:2020oko, Martinovic:2021hzy}, and so on. 
These methods work well to place upper limits on the inflationary SGWB. 
However, in these methods, it is impossible to guarantee that the remaining exotic component is of inflationary origin without a priori assumptions on an inflationary model as well as on the other cosmological sources. 
For example, although the slow-roll inflation predicts the spectral density $S_h(f) \propto f^{-\alpha}~(\alpha \simeq 3)$, 
it is not a universal prediction of inflation. 
Inflation can predict a wide variety of spectra, especially in models generating SGWB detectable by upcoming experiments \cite{Guzzetti:2016mkm, Bartolo:2016ami, Caprini:2018mtu}. 

The main purpose of this paper is to investigate whether the inflationary SGWB is distinguishable from the other components in direct-detection experiments without any a priori assumptions,  
focusing on a unique and universal prediction of inflation: the generation of superhorizon modes. 
\footnote{Several alternatives of inflation have been proposed to generate the superhorizon modes (see, e.g., Sec.~6.5 in Ref.~\cite{Caprini:2018mtu}). 
To be exact, our argument is also applied to these scenarios.}
Superhorizon modes are generated by inflation but not by any causal mechanism in the post-inflationary universe. 
In consequence, the inflationary GWs have a standing-wave nature after the horizon re-entry \cite{Grishchuk:1993ds,Allen:1999xw,Gubitosi:2017zoj,Gubitosi:2017zbz,Contaldi:2018akz}. 
As reviewed in section \ref{sec:flrw}, 
the standing-wave nature is observed as unusual properties of SGWB, 
most notably, correlations between GWs from opposite directions. 
Although this property has been already noticed in the literature, we would like to emphasize it as a unique prediction of inflation and name it {\it antipodal correlations}. 
The astrophysical SGWB, or any type of SGWB from localized sources, will not have such correlations because GWs from distant sources are uncorrelated with each other. 
Therefore, it is a unique signature for the inflationary SGWB. 

About 20 years ago, however, 
Allen et al. \cite{Allen:1999xw} showed that the unusual properties of the inflationary SWGB above cannot be detected in the stain correlation analysis. 
The antipodal correlations rapidly oscillate due to interference between GWs from opposite directions. 
It is inevitably smoothed out by averaging over frequencies unresolvable because of the finite observation time. 
Moreover, it was recently pointed out in Margalit et al. \cite{Margalit:2020sxp} that metric perturbations along the line-of-sight randomize the GW phases. 
This effect reduces the detectability of the antipodal correlations because the observed quantity is the strain smoothed over the sky with the finite angular resolution of a detector.  
The above two effects have been also pointed out for the three-point correlation function in Refs.~\cite{Bartolo:2018evs, Bartolo:2018rku, Bartolo:2019oiq}. 
As noted in Refs.~\cite{Margalit:2020sxp,Bartolo:2019oiq}, 
we need to use phase-incoherent methods such as the intensity map \cite{Renzini:2018vkx, Renzini:2018nee} to avoid these problems of interference. 
In this paper, we thus investigate whether we can construct a phase-incoherent estimator of the intensity map to detect the antipodal correlations. 
We found that the conclusion depends on whether the inflationary GWs have statistical isotropy or not. 
In the standard inflationary models with statistical homogeneity and isotropy, there is no estimator that is sensitive to the antipodal correlations but does not suffer from the problematic phase factor. 
On the other hand, it is possible to find a non-vanishing estimator of the antipodal correlations for inflationary models with statistical anisotropy. 
SGWB from {\it anisotropic} inflation is distinguishable from the other components. 

This paper is organized as follows.
In section \ref{sec:sb}, we briefly review possible properties of SGWB and define the antipodal correlation.
In section \ref{sec:sw}, we show the standing-wave nature of the inflationary GWs and how it leads to the antipodal correlations in SGWB. 
In section \ref{sec:detect}, after reviewing the detectability of the antipodal correlations in the strain correlation approach, we consider the intensity correlation approach.
Our conclusions are summarized in \ref{sec:conclusion}. 
In Appendix \ref{sec:time}, we also discuss the detectability of the antipodal correlations in the time domain analysis.

\section{\label{sec:sb} Stochastic Gravitational Wave Background}

In this section, 
we shortly review how SGWB can be characterized with emphasis on the unusual statistical properties of the inflationary SGWB.

The stochastic gravitational wave background is defined by a superposition of GWs from all directions of the sky. 
In the transverse-traceless gauge, 
it can be expanded as\footnote{In this paper, we denote a stochastic quantity with a hat.}
    \begin{align}
        &\hat{h}_{ij}(t,\vecx) = \nm \\
        &\quad \sum_{A} \int_{-\infty}^\infty {\rm d}f
        \int {\rm d}^2\hn \ 
        \hat{h}_A(f,\hn) e_{ij}^A(\hn) 
        e^{-2\pi i f(t-\hn \cdot \vecx)} \,,
    \label{eq:pwexpansion}
    \end{align}
in terms of plane waves with a frequency $f$ and a propagating direction $\hn$.
\footnote{We follow the notations in Maggiore's book \cite{Maggiore:2007ulw}. In some literature, $\hn$ is used for a direction on the sky, which is opposite to the propagating direction.}
The tensors $e_{ij}^A(\hn)$ are the polarization tensors for the two GW polarization states with normalization $e_{ij}^A(\hn)e^{B,ij}(\hn)=2\delta^{AB}$. 
To simplify the expressions below, 
we specialize to a circular polarization basis $e_{ij}^{\pm}(\hn)$: they are related to the ``plus-cross" polarization vectors as $e_{ij}^{\pm}(\hn) = e_{ij}^{\text{P.}}(\hn) \pm i \, e_{ij}^{\text{C.}}(\hn)$ and thus satisfy $e_{ij}^{\pm}(-\hn) = e_{ij}^{\mp}(\hn)$.
We define $t=0$ as the start time of observation. 

The Fourier amplitudes $\hat{h}_A(f,\hn)$ are random variables and their statistical distribution characterizes the stochastic background. 
Usually, we make the following assumptions on the statistical distribution:

\begin{enumerate}
    \item Gaussianity: \label{enu:gauss}
    all the statistical information in SGWB can be characterized by the two-point correlation function 
    $\langle \hat{h}^\dagger_A(f_1,\hn_1) \hat{h}_B(f_2, \hn_2) \rangle$.
    
    \item Isotropy: \label{enu:iso}
    the correlation functions are invariant under the rotation on the celestial sphere, i.e., 
    
    $\langle \hat{h}^\dagger_A(f_1,\hn_1) \hat{h}_B(f_2, \hn_2) \rangle$ depends on  $\hn_1$ and $\hn_2$ only through $\hn_1 \cdot \hn_2$. 
    
    \item No angular correlations \footnote{Note that the statistical isotropy \ref{enu:iso} does not forbid the angular correlations as is the case with the temperature map of cosmic microwave background (CMB). 
The properties \ref{enu:iso} and \ref{enu:no ang} are independent assumptions.} : \label{enu:no ang}
    GWs from different directions are not correlated with each other, i.e., 
    
     $\langle \hat{h}^\dagger_A(f_1,\hn_1) \hat{h}_B(f_2, \hn_2) \rangle \propto \delta^2(\hn_1,\hn_2)$. 
     
    \item Stationarity: \label{enu:stat}
    the correlation functions are invariant under the time translation, i.e., 
    
    $\langle \hat{h}^\dagger_A(f_1,\hn_1) \hat{h}_B(f_2, \hn_2) \rangle \propto \delta(f_1-f_2)$.
    
    \item Unpolarized:  \label{enu:pol}
    different polarization modes are independent and have the same statistics, i.e., 
    
    $\langle \hat{h}^\dagger_A(f_1,\hn_1) \hat{h}_B(f_2, \hn_2) \rangle \propto \delta_{AB}$,
    
    and the coefficient is independent of the polarizations. 
    
\end{enumerate}

When all these assumptions are satisfied, SGWB is characterized as
    \begin{align}{\label{eq:spect den}}
        &\langle \hat{h}^\dagger_A(f_1,\hn_1) \hat{h}_B(f_2, \hn_2) \rangle = \nm \\
        &\hspace*{2cm} \frac{S_h^{\text{(D)}}(f_1)\delta_{AB}}{4\pi} \delta(f_1-f_2)\delta^2(\hn_1,\hn_2) \,,
    \end{align}
with a (double-sided) spectral density $S^{\text{(D)}}_h(f)$. 

As shown in Refs.~\cite{Grishchuk:1993ds,Allen:1999xw,Gubitosi:2017zoj,Gubitosi:2017zbz,Contaldi:2018akz}, 
the inflationary SGWB does not satisfy the two assumptions \ref{enu:no ang} and \ref{enu:stat} as well as the last assumption \ref{enu:pol}:\footnote{The second term in Eq.~(\ref{eq:correlator}) is proportional to $\delta_{AB}$ in Ref.~\cite{Allen:1999xw}. As we will show in the next section, it should be replaced by $\delta_{A(-B)}$.}
the correlation function has an additional component as
    \begin{align}\label{eq:correlator}
        &\langle \hat{h}^\dagger_A(f_1,\hn_1) \hat{h}_B(f_2, \hn_2) \rangle = \nm \\
        &\hspace*{1cm}\quad \frac{S_h^{\text{(D)}}(f_1)\delta_{AB}}{4\pi}\delta(f_1-f_2)\delta^2(\hn_1,\hn_2) \nm \\
        &\hspace*{1cm} + \frac{A_h^{\text{(D)}}(f_1)\delta_{A(-B)}}{4\pi} \delta(f_1+f_2)\delta^2(\hn_1,-\hn_2) \,.
    \end{align}
Here, we have defined $S^{\text{(D)}}_h(f)$ and $A_h^{\text{(D)}}(f)$ as double-sided quantities. 
The corresponding single-sided spectral densities are defined by
    \begin{align}
        S_h(f) &\equiv S_h^{\text{(D)}}(f) + S_h^{\text{(D)}}(-f) = 2S_h^{\text{(D)}}(f) \,, \label{eq:single sh} \\
        A_h(f) &\equiv A_h^{\text{(D)}}(f) + A_h^{\text{(D)}}(-f) = 2{\rm Re}[ A_h^{\text{(D)}}(f) ] \,, \label{eq:single ah}
    \end{align}
respectively. 
The second new term in Eq.~(\ref{eq:correlator}) shows that GWs from the opposite directions are correlated, i.e., it corresponds to the antipodal correlations. 
In the next section, we will show that inflation universally predicts such correlations. 

\section{\label{sec:sw} Antipodal correlations}

In this section, 
we show how the inflationary GWs cause the antipodal correlations, i.e., the correlations between GWs from the opposite directions, in the observed SGWB. 
Although most arguments in this section have been presented in the literature \cite{Grishchuk:1993ds,Allen:1999xw,Gubitosi:2017zoj,Gubitosi:2017zbz,Contaldi:2018akz}, 
we rederive them in terms of realizations instead of statistically-averaged quantities for the later arguments on the construction of the estimator in Sec.~\ref{sec:detect}. 

\subsection{\label{sec:pre} Traveling/Standing-wave nature of stochastic gravitational wave background}

The expansion (\ref{eq:pwexpansion}) can be derived from the Fourier transform of GWs (see, e.g., Sec.~1.2 of Ref.~\cite{Maggiore:2007ulw}):
    \begin{align}\label{eq:fourier}
        \hat{h}_{ij}(t,\vecx) = \sum_{A=\pm}\int \frac{{\rm d}^3 k}{(2\pi)^3}~\hat{h}_A(t,\veck)e_{ij}^A(\hk) e^{i\veck \cdot \vecx} \,,
    \end{align}
where $\hat{h}^\dagger_{\pm}(t,\veck) = \hat{h}_{\pm}(t,-\veck)$. 
The vector $\hk$ is the unit vector along $\veck$: $\hk \equiv \veck/|\veck|$. 
In the local universe, 
$\hat{h}_{ij}(t,\vecx)$ satisfies the wave equation $\square \hat{h}_{ij}=0$ in a good approximation 
and therefore $\hat{h}_A(t,\veck)$ can be expanded into the positive and negative frequency modes as
    \begin{align}\label{eq:sol flat}
        \hat{h}_A(t,\veck) = \ampp_A(\veck)e^{-ikt} + \ampn_A(\veck)e^{ikt} \,.
    \end{align}
Here, the coefficients $\ampp_A(\veck)$ and $\ampn_A(\veck)$ are the integration constants and satisfy $\ampn_{\pm}(\veck) = [\ampp_{\pm}(-\veck)]^\dagger$ from  
the reality condition of $\hat{h}_{ij}(t,\vecx)$. 
In Eq.~(\ref{eq:sol flat}), 
the first (second) term represents a plane wave moving along $\hn = \hk$ ($- \hk$) with the frequency $f=k/2\pi$ ($-k/2\pi$). 
Therefore, the amplitude $\hat{h}_A(t,\hn)$ in Eq.~(\ref{eq:pwexpansion}) is read as
    \begin{align}\label{eq:rel amp}
         \hat{h}_{\pm}(f,\hn) 
       =
       \begin{cases}
        f^2\ampp_{\pm}(2\pi |f| \hn) \quad \text{for}~f > 0 \,, \\[6pt]
        f^2\ampn_{\mp}(-2\pi |f| \hn) \quad \text{for}~f < 0 \,.
       \end{cases}
    \end{align}
Note that the relation $[\hat{h}_{\pm}(f,\hn)]^\dagger = \hat{h}_{\mp}(-f,\hn)$ is satisfied as expected from the reality condition of $\hat{h}_{ij}(t,\vecx)$. 

The coefficients $\ampp_A(\veck)$ and $\ampn_A(\veck)$ are determined by the initial conditions and characterize the GW sources.
When all of them are independent, 
the GW background (\ref{eq:pwexpansion}) is given by the superposition of independent traveling waves. 
This is expected for SGWB from localized sources.
However,  
this is not the only possibility even when the statistical homogeneity is assumed \cite{Gubitosi:2017zbz}: 
the statistical homogeneity forbids the correlations between $\ampp_A(\veck)\,, \ampn_A(\veck)$ with different values of $\veck$ but not those between $\ampp_A(\veck)$ and $\ampn_A(\veck)$ with the same value of $\veck$: 
    \begin{align}
       \left\langle \left[ \ampp_A(\veck) \right]^\dagger \ampn_A(\veck) \right\rangle \neq 0 \,.
    \end{align}
From Eq.~(\ref{eq:rel amp}), 
this leads to the correlation between GWs with opposite frequencies and directions, i.e., the $A_h$ term in Eq.~(\ref{eq:correlator}). 
\footnote{This argument shows that SGWB can only have the antipodal correlations as angular correlations when the statistical isotropy and homogeneity are assumed for $\hat{h}_{ij}(t,\vecx)$. 
It can be also shown that the polarization dependence is restricted by imposing the invariance under the rotation around $\hk$, 
for which the circular polarization basis is transformed as $e_{ij}^{\pm}(\hk)\to e^{\pm 2i\psi} e_{ij}^{\pm}(\hk)$.}
In the next subsection, we will show that $\ampp_A(\veck)$ and $\ampn_A(\veck)$ have almost the same magnitude for the inflationary GWs. 
This means that the inflationary GWs have a standing-wave nature \cite{Grishchuk:1993ds,Allen:1999xw,Gubitosi:2017zoj,Gubitosi:2017zbz,Contaldi:2018akz}. 

\subsection{\label{sec:flrw} Propagation in the homogeneous universe}

To make the basic idea clearer, let us consider the propagation of the inflationary GWs in an idealistic homogeneous universe,
    \begin{align}
        {\rm d}s^2 = a^2(\eta)\left[ -{\rm d}\eta^2 + (\delta_{ij} + h_{ij}){\rm d}x^i{\rm d}x^j \right] \,,
    \end{align}
where we have introduced the conformal time $\eta$ and the scale factor $a(\eta)$. 
The scale factor is normalized as $a(\eta_0)=1$ for the start time of observation $\eta=\eta_0$. 
Thus, the comoving wavenumber below can be identified with the physical wavenumber in Eq.~(\ref{eq:fourier}). 
Inhomogeneities of the universe have large effects on the GW phases \cite{Bartolo:2018evs, Bartolo:2018rku, Margalit:2020sxp}. 
However, this does not change our conclusion on the detectability in the next section \ref{sec:detect} as we will give comments there.

Inflation generates stochastic GWs from vacuum fluctuations. 
A remarkable point is that inflation can generate the {\it superhorizon modes} with $k\eta \ll 1$ whereas the other causal mechanism in the post-inflationary universe cannot. 
On superhorizon scales, the solutions of the evolution equation in the expanding universe
    \begin{align}
        \hat{h}_A'' + 2{\cal H} \hat{h}_A' + k^2 \hat{h}_A = 0 \,,
    \end{align}
are constituted by constant and decaying modes. 
Here, ${\cal H}$ is the conformal Hubble parameter: ${\cal H} \equiv a'/a$. 
The prime $'$ denotes the derivative with respect to the conformal time $\eta$. 
Shortly after the horizon crossing during inflation, 
the amplitude of the decaying mode decreases quickly and thus
the Fourier amplitude $\hat{h}_A(\eta, \veck)$ only contains a single statistical variable: 
	\begin{align}\label{eq:sol sh}
		\hat{h}_A(\eta, \veck) \to \chi_k(\eta)\hat{h}^{\text{(prim)}}_{A,\veck} \,,
	\end{align}
where $\chi(k)$ is the transfer function with $\chi_k(\eta) \to 1$ for $k\eta \ll 1$. 
In the standard inflationary scenario, 
the primordial amplitudes $\hat{h}^{\text{(prim)}}_{A,\veck}$ are Gaussian random variables with statistical homogeneity and isotropy:
    \begin{align}\label{eq:primordial p}
        \langle~ [ h^{(\text{prim})}_{A,\veck_1}]^{\dagger} \hat{h}^{\text{(prim)}}_{B,\veck_2} ~\rangle = \delta_{AB} P_h(k_1)\delta^3(\veck_1-\veck_2) \,.
    \end{align}
Matching the local solution (\ref{eq:sol flat}) to the superhorizon solution (\ref{eq:sol sh}), 
we find that the two amplitudes $\ampp_A(\veck)$, $\ampn_A(\veck)$ should be correlated with each other for the inflationary GWs.

We can estimate the correlations between the positive and negative frequency modes by solving the following evolution equation for the transfer function, 
    \begin{align}\label{eq:evolution eq}
        \chi_k'' + 2{\cal H} \chi_k' + k^2 \chi_k = 0 \,.
    \end{align}
In the subhorizon regime, it has the WKB solutions $\chi_k(\eta) \propto e^{\pm ik\eta}/a$. 
Imposing the initial condition $\chi_k(\eta) \to 1$ in the superhorizon regime, 
the subhorizon solution has both positive and negative frequency modes with the same amplitude because $\chi_k$ should be real: 
    \begin{align}\label{eq:sol hom0}
        \chi_k(\eta) = \frac{\alpha_k e^{-ik\eta} + \alpha_k^\ast e^{ik\eta}}{a(\eta)} \,,
    \end{align}
where $\alpha_k$ is a constant.
We can find an analytic solution,
    \begin{align}
        \chi_k(\eta) = \frac{e^{ik\eta} - e^{-ik\eta}}{2ik\eta} \,,
    \end{align}
in the radiation-dominated era, where the relevant modes for the GW interferometers re-enter the horizon. 
This solution can be rewritten as
    \begin{align}\label{eq:sol hom}
        \chi_k(\eta) = \frac{a(\eta_k)}{a(\eta)} \left(\frac{e^{ik\eta} - e^{-ik\eta}}{2i}\right) \,,
    \end{align}
introducing the horizon re-entry time $\eta_k$ by $k\eta_k = 1$.
The coefficients $\alpha_k$ in Eq.~(\ref{eq:sol hom0}) at the present time are obtained by connecting this solution to the late-time universe. 
Unless a nonadiabatic transition occurs, 
the solution at the present time is given in the form (\ref{eq:sol hom}) 
(see Refs. \cite{Watanabe:2006qe, Saikawa:2018rcs} for a more accurate transfer function).

Comparing Eq.~(\ref{eq:sol hom}) with Eq.~(\ref{eq:sol flat}), 
we find
    \begin{align}
    	\begin{aligned}\label{eq:apn hom} 
         \ampp_A(\veck) &= \overline{\cal T}_k e^{-i(k\eta_0-\frac{\pi}{2})} \hat{h}^{\text{(prim)}}_{A,\veck} \,,\\
         \ampn_A(\veck) &= \overline{\cal T}_k e^{i(k\eta_0-\frac{\pi}{2})} \hat{h}^{\text{(prim)}}_{A,\veck} \,,
        \end{aligned}
    \end{align}
with
    \begin{align}\label{eq:tk}
         \overline{\cal T}_k \equiv \frac{1}{2}\frac{a(\eta_k)}{a(\eta_0)} \,.
    \end{align}
Here, we have rewritten the conformal time $\eta$ in terms of the cosmic time $t$ as $\eta \simeq \eta_0 + t/a(\eta_0)$. 
\footnote{We have estimated the cosmic time as $t = \int_{\eta_0}^\eta a(\eta'){\rm d}\eta' \simeq a(\eta_0)(\eta-\eta_0)$ by neglecting the evolution of $a(\eta)$ during the observation.}
The damping factors and phase shifts in Eq.~(\ref{eq:apn hom}) 
are geometrically determined. 
As expected, the two amplitudes $\ampp_A(\veck)$, $\ampn_A(\veck)$ are represented by the single statistical variable $\hat{h}^{\text{(prim)}}_{A,\veck}$ and thus are correlated with each other.
Substituting these results to Eq.~(\ref{eq:rel amp}), 
we find
    \begin{align}\label{eq:rel amp prim}
         &\hat{h}_{\pm}(f,\hn) 
       = \nm \\
       & \quad
       \begin{cases}
        f^2 \overline{\cal T}_{2\pi|f|} e^{-2\pi i f \eta_0 + \frac{i\pi}{2}} \hat{h}^{\text{(prim)}}_{\pm,2\pi f\hn} \quad \text{for}~f > 0 \,, \\[6pt]
        f^2 \overline{\cal T}_{2\pi|f|} e^{-2\pi i f \eta_0 - \frac{i\pi}{2}} \hat{h}^{\text{(prim)}}_{\mp,2\pi f\hn} \quad \text{for}~f < 0 \,,
       \end{cases}
    \end{align}
and thus the following relation: 
    \begin{align}\label{eq:antipodal h}
         h_{\pm}(f,\hn) = -e^{4\pi if \eta_0} h_{\mp}(-f,-\hn) \,.
    \end{align}
This relation shows that:
\begin{enumerate}

\item[(i)] There is a one-to-one correspondence between {\it realizations} of SGWB with the opposite frequencies, directions, and circular polarizations. 

\item[(ii)] Their amplitudes are the same. 

\item[(iii)] Their phase difference is huge and proportional to the frequency $f$. 

\end{enumerate}
%

\begin{figure}[t]
    \centering
    \includegraphics[width=.95\linewidth]{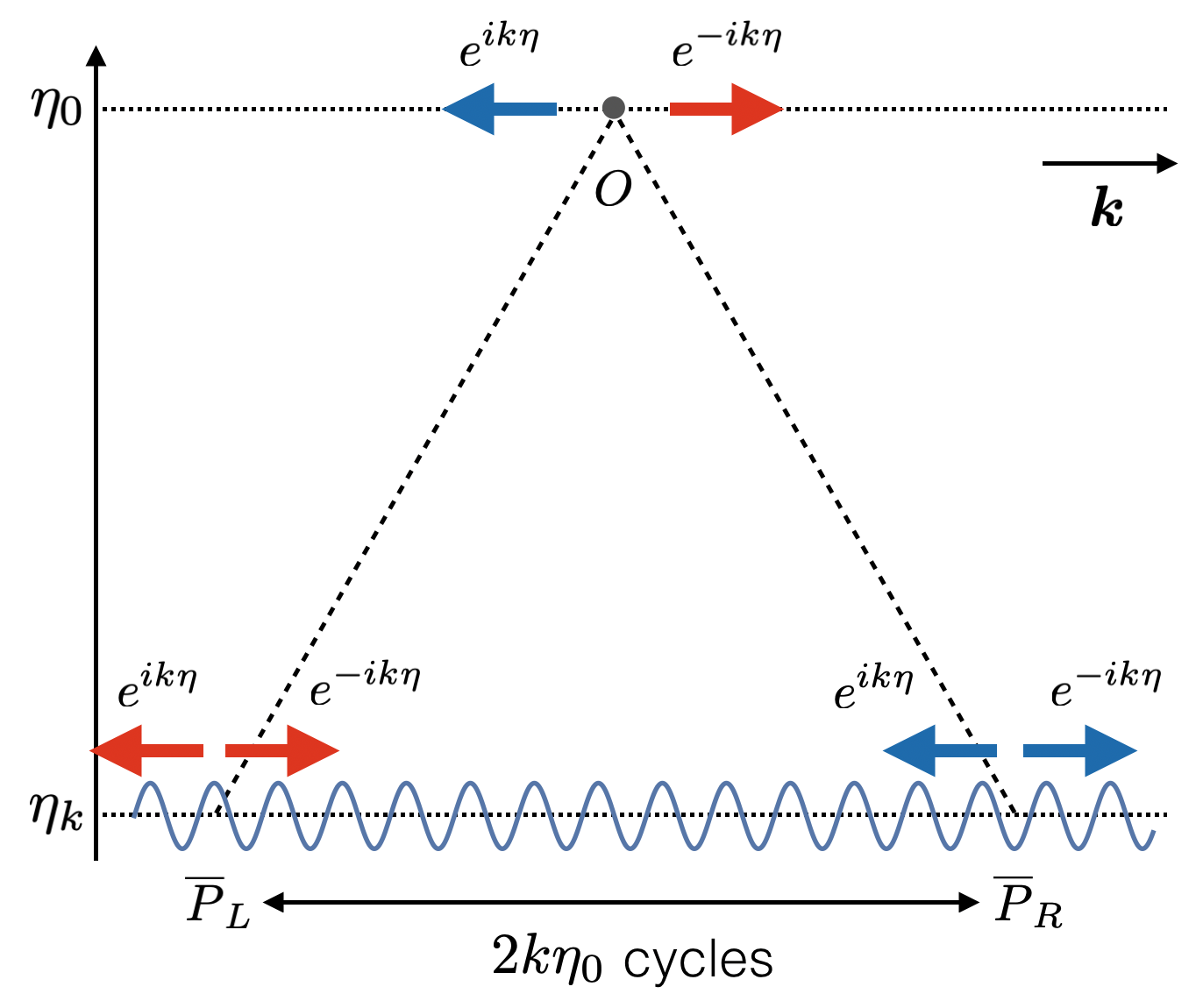}
    \caption{Propagation of the inflationary GWs in the homogeneous universe (vertical: the time direction, horizontal: the spatial direction parallel to $\veck$). The point $O$ represents the observer's position. The points $\overline{P}_L$ and $\overline{P}_R$ represent the points where right- and left-moving null geodesics cross the constant-time hypersurface $\eta = \eta_k$, respectively.}
    \label{fig:homogeneous}
\end{figure}

These results can be easily understood from Fig.~\ref{fig:homogeneous}. 
The inflationary GWs induce coherent standing waves on the constant-time hypersurface $\eta = \eta_k$; 
left- and right-moving modes are emitted with the same amplitude and the definite phase difference at each point. 
The positive (negative) frequency modes at the observer's position $O$ are the right (left) moving modes coming from the point $\overline{P}_L$ ($\overline{P}_R$). 
The amplitudes of the positive and negative frequency modes are the same because the right and left moving modes are damped by the cosmic expansion at the same rate. 
Since the phase is conserved along the null geodesic, 
the phase difference is given by the number of cycles between $\overline{P}_L$ and $\overline{P}_R$, $2k\eta_0$, with a small correction from the intrinsic phase difference between the right- and left-moving modes on the hypersurface $\eta = \eta_k$. 

Using Eq.~(\ref{eq:antipodal h}), 
it is easy to find the relation
    \begin{align}\label{eq:ah sh}
         A_h^{\text{(D)}}(f) = -S_h^{\text{(D)}}(f)e^{4\pi i f \eta_0} \,,
    \end{align}
between the two spectral densities in Eq.~(\ref{eq:correlator}).
Therefore, inflation predicts large antipodal correlations.

\section{\label{sec:detect} (Un)Detectability of the antipodal correlations}

\subsection{\label{sec:allen} The argument in Allen et al. (1999)}

We review the argument in Allen et al. (1999) \cite{Allen:1999xw} on the undetectability of the $A_h$ term in Eq.~(\ref{eq:correlator}). 
The $A_h$ term in Eq.~(\ref{eq:ah sh}) is a highly oscillating function of $f$. 
Its period is of the order of $1/T_{\rm age}$ for the age of the Universe $T_{\rm age} \sim \eta_0$. 
This oscillation has a clear physical interpretation: it is interference between GWs from the antipodal points $\overline{P}_L$ and $\overline{P}_R$ in Fig.~\ref{fig:homogeneous}. 

The point of the argument is that $\hat{h}_A(f,\hn)$ is {\it not} an observable: 
our frequency resolution is fundamentally limited by the observation time $T$ as $\Delta f \sim 1/T \gg 1/T_{\rm age}$. 
To take into account the finite frequency resolution,  
we introduce the smoothed quantity,
    \begin{align}\label{eq:smoothed h}
       \hat{h}_{A;T}(f,\hn) \equiv \int_{-\infty}^{\infty} \!{\rm d}f'~ W_T(f-f') \hat{h}_A(f',\hn) \,,
    \end{align}
and consider it as observable. 
Here, $W_T(f)$ is the window function with the width $\Delta f \sim 1/T$. 
For example, when we use the short-time Fourier transform, 
    \begin{align}\label{eq:ft fourier}
       \hat{h}_{A;T}(f,\hn) = \int_0^{T}\!{\rm d}t~ \hat{h}_A(t,\hn)e^{2\pi i f t} \,,
    \end{align}
the window function is given by
    \begin{align}\label{eq:window}
        W_T(f) =  \frac{e^{i\pi f T} \sin(\pi f T)}{\pi f} \,.
    \end{align}
Computing the antipodal correlations for the smoothed quantity (\ref{eq:smoothed h}), 
we find
    \begin{align}\label{eq:smoothed ah}
       &\langle \hat{h}^\dagger_{\pm;T}(f,\hn)  \hat{h}_{\mp;T}(-f,-\hn) \rangle = \nm \\
       &\hspace*{2cm} \int_{-\infty}^{\infty} \!{\rm d}f'~ |W_T(f-f')|^2 A_h^{\text{(D)}}(f') \,,
    \end{align}
and thus
    \begin{align}
       &\langle \hat{h}^\dagger_{\pm;T}(f,\hn)  \hat{h}_{\mp;T}(-f,-\hn) \rangle = \nm \\
       &\hspace*{1cm} -\int_{-\infty}^{\infty} \!{\rm d}f'~ |W_T(f-f')|^2 S_h^{\text{(D)}}(f')e^{4\pi i f'\eta_0} \,,
    \end{align}
using the relation (\ref{eq:ah sh}). 
Therefore, even when we take the best resolution $\Delta f \sim 1/T$, 
the $A_h$ term is erased by smoothing over the unresolvable frequencies in eq.~(\ref{eq:smoothed ah}) 
unless the spectral density $S_h^{\text{(D)}}(f)$ has a very sharp peak with a width much less than $1/T_{\rm age}$.  
The situation becomes worse when we take into account the inhomogeneities. 
The inhomogeneities introduce the $\hn$-dependent phase $e^{2\pi i f \eta_0 \hat{\phi}(\hn)}$ in eq.~(\ref{eq:ah sh}) with the function $\hat{\phi}(\hn)$ written in terms of the gravitational potential along the line-of-sight \cite{Bartolo:2018evs, Bartolo:2018rku, Margalit:2020sxp}. 
Thus, the $A_h$ term also rapidly oscillates for the direction $\hn$ and vanishes when smoothed over $\hn$. 
In conclusion, 
provided that the spectral density $S_h(f)$ slowly varies with respect to $f$, 
the correlation function for the smoothed field becomes
    \begin{align}\label{eq:smoothed h 2pt}
       &\langle \hat{h}^\dagger_{A;T}(f_1,\hn_1)  \hat{h}_{B;T}(f_2,\hn_2) \rangle = \nm \\
       &\hspace*{2cm} \frac{S_{h}^{\text{(D)}}(f_1)\delta_{AB}}{8\pi}\delta_T(f_1-f_2)\delta^2(\hn_1,\hn_2) \,,
    \end{align}
where
    \begin{align}\label{eq:deltaT}
        \delta_T(f_1-f_2) \equiv \int_{-\infty}^{\infty} \!{\rm d}f'~ W_T(f_1-f_2+f')W^\ast_T(f') \,,
    \end{align}
and indistinguishable from a non-inflationary SGWB (\ref{eq:spect den}). 

The root of the cancellation is the fact that the phase difference between $h_{\pm}(f,\hn)$ and $h_{\mp}(-f,-\hn)$ is rapidly oscillating with respect to the frequency $f$ (see Eq.~(\ref{eq:antipodal h})). 
This motivates us to use the intensity map,
    \begin{align}
       \hat{I}_A(f,\hn) \equiv |\hat{h}_A(f,\hn)|^2 \,,
    \end{align}
to detect the antipodal correlations. 
From Eq.~(\ref{eq:antipodal h}), 
it is easy to see that there is a coincidence between the {\it realizations} of the intensity with 
the opposite 
directions:
    \begin{align}\label{eq:antipodal i}
       \hat{I}_\pm(f,\hn) = \hat{I}_\pm(f,-\hn) \,,
    \end{align}
by using the reality condition $[\hat{h}_{\pm}(f,\hn)]^\dagger = \hat{h}_{\mp}(-f,\hn)$. 
This relation is not modified much even when we take into account the propagation through the inhomogeneous universe, because the modification is the order of the cosmological perturbations \cite{Laguna:2009re}. 
Since there is no problematic phase factor in Eq.~(\ref{eq:antipodal i}), 
the intensity map would work to detect the antipodal correlations. 
In the next subsection, 
we will discuss this possibility. 

\subsection{\label{sec:intensity} Antipodal correlations in the intensity map}

In this subsection, 
we discuss whether the antipodal correlations can be detected by using the intensity map,
    \begin{align}\label{eq:intensity}
       \hat{I}_A(f,\hn) \equiv |\hat{h}_A(f,\hn)|^2 \,.
    \end{align}
To take into account the finite frequency resolution, 
we introduce the intensity of the smoothed quantity (\ref{eq:smoothed h}) by
    \begin{align}\label{eq:smoothed i}
       \hat{I}_{A;T}(f,\hn) \equiv |\hat{h}_{A;T}(f,\hn)|^2 \,,
    \end{align}
and investigate whether the antipodal relation (\ref{eq:antipodal i}) can be confirmed through it. 
We would like to remark that
the quantity (\ref{eq:smoothed i}) is {\it not} the smoothing of the intensity (\ref{eq:intensity}):
    \begin{align}
       \hat{I}_{A;T}(f,\hn) \neq \int_{-\infty}^{\infty} \!{\rm d}f' |W_T(f-f')|^2 \hat{I}_A(f',\hn) \,,
    \end{align}
while it is true when the ensemble average is taken:
    \begin{align}
       \langle \hat{I}_{A;T}(f,\hn) \rangle = \int_{-\infty}^{\infty} \!{\rm d}f' |W_T(f-f')|^2 \langle \hat{I}_A(f',\hn) \rangle \,.
    \end{align}
By using the relations (\ref{eq:rel amp prim}),
we can rewrite the smoothed intensity (\ref{eq:smoothed i}) as
\begin{widetext}
    \begin{align}\label{eq:smoothed i expansion}
        &\hat{I}_{A;T}(f,\hn) = \nm \\
        &\qquad \int_{-\infty}^{\infty} \!{\rm d}f'\int_{-\infty}^{\infty} \!{\rm d}f''~  W^\ast_T(f-f') W_T(f-f'') f'^2f''^2 \overline{\cal T}_{2\pi |f'|}\overline{\cal T}_{2\pi |f''|}~ e^{-2\pi i(f''-f')\eta_0} [\hat{h}^{\text{(prim)}}_{A',2\pi f' \hn}]^\dagger \hat{h}^{\text{(prim)}}_{A'',2\pi f'' \hn} \,,
    \end{align}
\end{widetext}
where $A'$ and $A''$ are $+A$ for $f', f''>0$ and $-A$ for $f', f''<0$. 
We can see that 
the problematic phase factor $e^{-2\pi i(f''-f')\eta_0}$ in the smoothed intensity $\hat{I}_{A;T}(f,\hn)$ remains unless the other factor in the integrand has a sharp peak at $f''=f'$ with the width $|f''-f'| \ll 1/T_{\rm age}$. 
This is not the case for the realization (\ref{eq:smoothed i}). 
Therefore, 
unlike the unsmoothed intensity (\ref{eq:intensity}), 
the antipodal relation (\ref{eq:antipodal i}) does not hold for the realizations of the smoothed intensity $\hat{I}_{A;T}(f,\hn)$: 
    \begin{align}\label{eq:antipodal smoothed i}
       \hat{I}_{\pm;T}(f,\hn) \neq \hat{I}_{\pm;T}(f,-\hn) \,.
    \end{align}
The different phase factor $e^{+2\pi i(f''-f')\eta_0}$ appears for $\hat{I}_{\pm;T}(f,-\hn)$ instead of $e^{-2\pi i(f''-f')\eta_0}$ in Eq.~(\ref{eq:smoothed i expansion}).

Here, we discuss whether it is possible to test the antipodal relation (\ref{eq:antipodal i}) by constructing an appropriate estimator.
First, we can see that the higher-order statistics is of no use for this purpose. 
This becomes clear by decomposing the smoothed intensity $\hat{I}_{A;T}(f,\hn)$ into the ensemble average and the deviation from it:
   \begin{align}\label{eq:del int}
		\hat{I}_{A;T} = \langle \hat{I}_{A;T} \rangle + \delta \hat{I}_{A;T} \,.
	\end{align}
These two terms $\langle \hat{I}_{A;T} \rangle$ and $\delta \hat{I}_{A;T}$ correspond to the contributions from $f'' = f'$ and $f'' \neq f'$ respectively in the integral (\ref{eq:smoothed i expansion}) because $\langle [\hat{h}^{\text{(prim)}}_{A,2\pi f' \hn}]^\dagger \hat{h}^{\text{(prim)}}_{A,2\pi f'' \hn} \rangle$ contains $\delta(f''-f')$. 
Therefore, the problematic phase factor remains in the deviation $\delta \hat{I}_{A;T}$ and spoils the antipodal relation. 
In fact, 
we can show that the antipodal contribution in the two-point function vanishes with assuming the Gaussianity of $\hat{h}_A(f,\hn)$: 
    \begin{align}\label{eq:smoothed i 2pt}
      &C_I(\hn_1, \hn_2) \equiv \langle \delta \hat{I}_{A;T}(f_1,\hn_1) \delta \hat{I}_{B;T}(f_2,\hn_2) \rangle \,,
    \end{align}
can be rewritten in terms of the correlation functions of $\hat{h}_{A;T}(f,\hn)$ as
    \begin{align}\label{eq:wick}
      &C_I(\hn_1, \hn_2) = |\langle \hat{h}^\dagger_{A;T}(-f_1,\hn_1)\hat{h}_{B;T}(f_2,\hn_2) \rangle|^2 \nm \\
      &\hspace*{2cm} + |\langle \hat{h}^{\dagger}_{A;T}(f_1,\hn_1)\hat{h}_{B;T}(f_2,\hn_2) \rangle|^2 \,,
    \end{align}
where the reality condition $\hat{h}^\dagger_{A;T}(f,\hn)=\hat{h}_{A;T}(-f,\hn)$ has been used. 
Using the expression (\ref{eq:smoothed h 2pt}) for the correlation functions of $\hat{h}_{A;T}(f,\hn)$, 
we can find
    \begin{align}
        C _I(\hn_1, \hn_2) \propto \delta^2(\hn_1,\hn_2) \,,
    \end{align}
and the coefficient is written only in terms of the spectral density $S_h(f)$. 
From similar arguments, we can show that higher-point functions are of no use for testing the antipodal relation (\ref{eq:antipodal i}).

\begin{figure}[t]
    \centering
    \includegraphics[width=.95\linewidth]{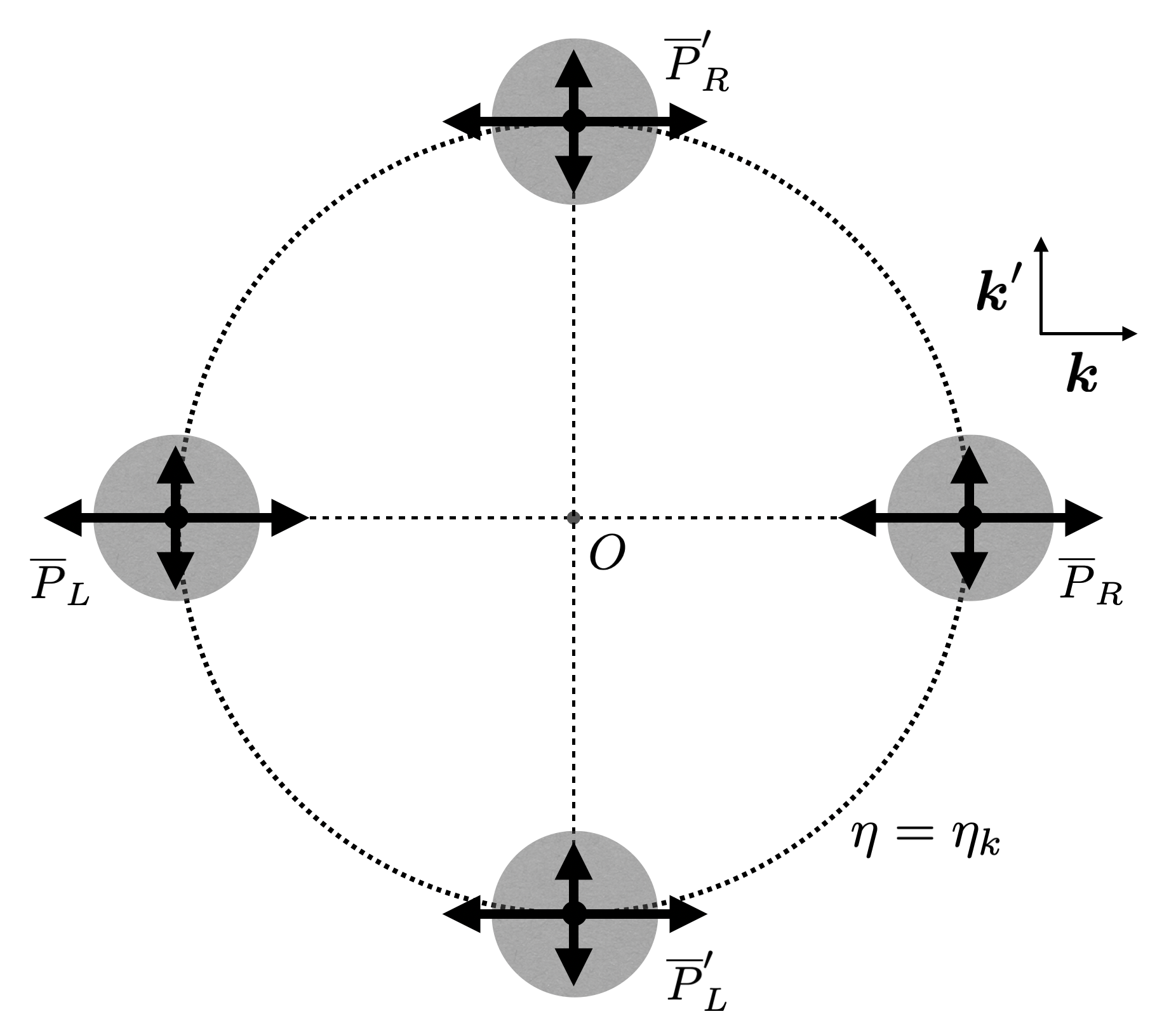}
    \caption{The angular correlations in the intensity map. The circle of the dotted line represents the intersection between the past light cone and the hypersurface $\eta=\eta_k$. The direction and length of the arrows indicate the moving direction and intensity of GWs, respectively. The intensity of GWs is the same for modes with parallel moving directions and uncorrelated between modes with nonparallel moving directions (see Eq.~(\ref{eq:primordial p})). 
    The radius of the shaded circles represents the expectation values of the intensity. 
    It is independent of the moving direction and the emission point in the standard inflationary models with statistical homogeneity and isotropy as depicted in the figure. 
    }
    \label{fig:intensity_map}
\end{figure}

The remaining possibility is a one-point function. 
The problematic phase factor in Eq.~(\ref{eq:smoothed i expansion}) is erased in the ensemble average
    \begin{align}\label{eq:average i}
       \langle \hat{I}_{A;T}(f,\hn) \rangle \equiv I_{A;T}(f, \hn) \,,
    \end{align}
because $\langle [\hat{h}^{\text{(prim)}}_{A,2\pi f' \hn}]^\dagger \hat{h}^{\text{(prim)}}_{A,2\pi f'' \hn} \rangle$ contains $\delta(f''-f')$. 
This quantity is used for mapping SGWB in the literature ({\it e.g.,} Refs.~\cite{Mitra:2007mc, Renzini:2018vkx}). 
However, 
the averaging simultaneously erases the directional dependence in the intensity when the statistical isotropy is assumed: introducing the anisotropies
    \begin{align}\label{eq:anisotropies i}
        \Delta I_{A;T}(f, \hn) \equiv I_{A;T}(f, \hn) - \bar{I}_{A;T}(f) \,, 
    \end{align}
with the angular average in the sky $\bar{I}_{A;T}(f)$,
    \begin{align}\label{eq:del int av}
        \Delta I_{A;T}(f, \hn) = 0 \,.
    \end{align}
Therefore, 
we cannot find an estimator of the intensity that is sensitive to the antipodal correlations but does not suffer from the problematic phase factor in the standard inflationary modes with the statistical homogeneity and isotropy (\ref{eq:primordial p}). 
We have illustrated the situation in Fig.~\ref{fig:intensity_map}. 

%
%

The situation changes for inflationary models with statistical anisotropy, {\it i.e.,} hypothesis (b) in the section \ref{sec:sb} is broken (see, {\it e.g.},  Refs.~\cite{Kanno:2008gn, Watanabe:2009ct, Kanno:2010nr, Soda:2012zm} for concrete models)
    \begin{align}\label{eq:primordial p aniso}
        \langle~ [ h^{(\text{prim})}_{A,\veck_1}]^{\dagger} \hat{h}^{\text{(prim)}}_{B,\veck_2} ~\rangle = \delta_{AB} P_h(\veck_1)\delta^3(\veck_1-\veck_2) \,.
    \end{align}
In this case, the anisotropies in the averaged intensity are not erased, 
    \begin{align}
        \Delta I_{A;T}(f, \hn) \neq 0 \,,
    \end{align}
while the sharp peak $\delta(f'-f'')$ still appears due to the statistical homogeneity 
and thus the problematic phase factor disappears:
	\begin{align}\label{eq:int av aniso}
		&I_{A;T}(f,\hn) = \nm \\
		&
		\qquad \int_{-\infty}^{\infty} \!{\rm d}f'~  |W_T(f-f')|^2 f'^4 \overline{\cal T}_{2\pi |f'|}^2 P_h(2\pi f' \hn) \,.
	\end{align}
The intensity (\ref{eq:int av aniso}) satisfies the antipodal relation for the anisotropies
	\begin{align}\label{eq:antipodal deli}
		\Delta I_{\pm;T}(f, \hn)  = \Delta I_{\pm;T}(f, -\hn) \,,
	\end{align}
as a consequence of the standing-wave nature of the inflationary GWs (\ref{eq:apn hom}). 
Therefore, the inflationary SGWB can be distinguished from the other components if we detect (i) non-vanishing anisotropies $\Delta I_{A;T}(f, \hn)$ and (ii) their antipodal relation (\ref{eq:antipodal deli}). 

Let us also comment on the case when hypothesis (e) on polarization is broken \cite{Lue:1998mq, Contaldi:2008yz, Takahashi:2009wc, Sorbo:2011rz, Maleknejad:2011sq, Anber:2012du, Adshead:2013qp, Dimastrogiovanni:2016fuu}. 
In this case, we can show antipodal relations for all the Stokes parameters through the relation (\ref{eq:rel amp prim}). 
With the same arguments above, these antipodal relations are undetectable in the isotropic case and detectable in the anisotropic case. In the detectable case, they will give more evidence for the inflationary GWs. 

Before closing this section, 
it might be noteworthy to mention a difference from CMB. 
In contrast to GWs, electromagnetic waves (EMWs) are scattered many times by electrons in the early universe. 
Therefore, the angular correlations intrinsic in EMWs are erased, and there is no counterpart of the antipodal correlations in the CMB anisotropies. 
Instead, the CMB angular correlations are a tracer of the inhomogeneous background: 
the intensity is spatially modulated in the vicinity of an emission point by long-wavelength perturbations, 
and EMWs in these regions are scattered into the line-of-sight direction. 
The statistical isotropy at each emission point is locally broken by the long-wavelength perturbations. 
Moreover, 
the long-wavelength perturbations also break the statistical homogeneity among the emission points in Fig.~\ref{fig:intensity_map}. 
Therefore, the argument of Eq.~(\ref{eq:del int av}) is not applied to this type of angular correlation. 

\section{\label{sec:conclusion} Conclusion} 

The measurement of the inflationary SGWB is one of the main goals of future GW experiments. 
One obstacle to achieving it is the isolation of the inflationary SGWB from the other components generated by the unresolvable astronomical and cosmological GW sources. 
In this paper, we argued the detectability of a unique and universal property of the inflationary SGWB: antipodal correlations, i.e., correlations of GWs from opposite directions. 

It was argued in Allen et al. \cite{Allen:1999xw} that the conclusion is negative when we use a phase-coherent method, i.e., the standard strain correlation analysis, due to the phase oscillation unresolvable in the observation time. 
We thus investigate whether we can construct a phase-incoherent estimator of the intensity map to detect the antipodal correlations. 
We found that the conclusion depends on whether the inflationary GWs have statistical isotropy or not. 
Under the standard assumption of statistical homogeneity and isotropy, 
it is impossible to find an observable that is sensitive to the antipodal correlations but does not suffer from the problematic phase factor: 
the intensity constructed from the observed GW strain still has the annoying phase factor that erases the antipodal correlations. 
The ensemble average can get rid of the phase factor but simultaneously drops the angular information due to statistical isotropy. 
However, the latter argument is not applied to the inflationary models with statistical anisotropy. 
We can find a non-vanishing observable for the antipodal correlations and thus conclude that SGWB from {\it anisotropic} inflation is distinguishable from the other components. 

Our argument can be applied to other types of angular correlations. 
The problematic phase factor erases any types of angular correlations in the strain and non-averaged intensity. 
The ensemble average can get rid of the problematic phase factor but simultaneously drops the angular information under statistical isotropy. 
On the other hand, we can measure the angular correlations in CMB even under statistical isotropy. 
A natural question is thus whether we can find a way to measure the angular correlations in SGWB \footnote{The Boltzmann approach \cite{Contaldi:2016koz} is widely used to estimate the SGWB anisotropies as well as the CMB anisotropies. 
Our argument implies that we need to carefully discuss how an estimator of the distribution function (intensity, energy density) should be defined. }
and what kind of angular correlations are detectable. 
As we have remarked in Sec.~\ref{sec:intensity}, 
the local violation of statistical isotropy and homogeneity by long-wavelength perturbations is crucial for the detectability of the CMB angular correlations. 
Because the CMB angular correlations are detectable, it would be possible to find an estimator for the SGWB angular correlations induced by long-wavelength perturbations, e.g., through propagation and long-short wavelength mode couplings \cite{Laguna:2009re, Alba:2015cms, Dimastrogiovanni:2019bfl, Dimastrogiovanni:2021mfs}. 
In a subsequent paper, 
we will discuss how we should define the estimator to get rid of the problematic phase factor with (partially) keeping the angular information.

\begin{acknowledgments}
We thank the anonymous referee for the helpful suggestion to discuss the anisotropic case. 
This research was supported by the JSPS Grant-in-Aid for Scientific Research  (No.~17K14286, No.~19H01891, No.~20H05860) and JST SPRING (No.~JPMJSP2111).
\end{acknowledgments}

\appendix

\section{\label{sec:time} Correlation analysis in the time domain} 

In the main text, we have shown that the antipodal correlations cannot be detected with the maps of the Fourier amplitude. 
In both methods, the root of the undetectability is the fundamental limitation in frequency resolution due to the finite observation time. 
In this Appendix, we will show the same fact for the original signal (\ref{eq:pwexpansion}) without taking its finite-time Fourier transform (\ref{eq:ft fourier}) to confirm that the undetectability discussed in section \ref{sec:allen} is not a result of the limitation of the Fourier analysis. 

We compute the following correlation functions in the time domain:
    \begin{align}\label{eq:correlator time}
        \langle \hat{h}_A(t-\tau/2,\hn_1) \hat{h}_B(t+\tau/2,\hn_2) \rangle \,,
    \end{align}
where
    \begin{align}
        &\hat{h}_A(t,\hn) \equiv
        \int_{-\infty}^\infty {\rm d}f \
        \hat{h}_A(f,\hn)
        e^{-2\pi i f t} \,.
    \end{align}
The correlation functions of the filtered signals and the intensity map can be written in terms of them. 

For the standard contribution $\hn_1=\hn_2 (=\hn)$, 
the correlation functions (\ref{eq:correlator time}) are computed as
    \begin{align}
        &\langle \hat{h}_A(t-\tau/2,\hn) \hat{h}_B(t+\tau/2,\hn) \rangle \nm \\
        &\hspace*{2cm} =  \delta_{AB}\int_{-\infty}^\infty {\rm d}f \
        S^{\text{(D)}}_h(f) e^{2\pi i f \tau} \,.
    \end{align}
The result is independent of $t$. 
Therefore, we can use $\hat{h}_A(t-\tau/2,\hn) \hat{h}_B(t+\tau/2,\hn)$ for different values of $t$ as samples to estimate
    \begin{align}\label{eq:cs}
        C_S(\tau) \equiv \int_{-\infty}^\infty {\rm d}f \
        S^{\text{(D)}}_h(f) e^{2\pi i f \tau} \,.
    \end{align}
The function $C_S(\tau)$ is not small for sufficiently small values of $\tau$ because
    \begin{align}\label{eq:cs0}
        C_S(0) = \int_{-\infty}^\infty {\rm d}f \
        S^{\text{(D)}}_h(f) \,,
    \end{align}
and the spectral density $S^{\text{(D)}}_h(f)$ is positive semi-definite. 
We can estimate the spectral density $S^{\text{(D)}}_h(f)$ by taking the short-time Fourier transform of $C_S(\tau)$. 
This corresponds to the fact shown in the previous section. 
To further increase the sensitivity, 
in the standard correlation analysis, 
we usually apply the optimal filter $Q_S(\tau)$ to the estimator of $C_S(\tau)$ assuming the shape of $S^{\text{(D)}}_h(f)$ (e.g. Ref.~\cite{Maggiore:2007ulw}): 
    \begin{align}\label{eq:signal s filter}
        &\int_0^T{\rm d}\tau \ C_S(\tau)Q_S(\tau) =
        \int_{-\infty}^\infty {\rm d}f \
        S^{\text{(D)}}_h(f) Q_{S;T}(f) \,,
    \end{align}
where $Q_{S;T}(f)$ is the short-time Fourier transform of $Q_S(\tau)$. 
\footnote{To be exact, the integration domain for $\tau$ is not $[0,T]$. However, we do not need to care about it because the optimal filter $Q_A(\tau)$ decays quickly as $\tau$ increases.}
The positive semi-definiteness of $S^{\text{(D)}}_h(f)$ ensures that the signal (\ref{eq:signal s filter}) for the overall amplitude of the spectrum can be enhanced compared to the noise by choosing the filter function $Q_S(\tau)$ appropriately.

For the antipodal contribution $\hn_1=-\hn_2 (=\hn)$, 
the correlation functions (\ref{eq:correlator time}) are computed as
    \begin{align}
        &\langle \hat{h}_A(t-\tau/2,\hn) \hat{h}_B(t+\tau/2,-\hn) \rangle \nm \\
        &\hspace*{2cm} =  \delta_{A(-B)}\int_{-\infty}^\infty {\rm d}f \
        A^{\text{(D)}}_h(f) e^{-4\pi i f t} \,.
    \end{align}
The result is independent of $\tau$. Therefore, we can use $\hat{h}_A(t-\tau/2,\hn) \hat{h}_B(t+\tau/2,-\hn)$ for different values of $\tau$ as samples to estimate
\footnote{Note that the subscript $A$ of $C_A$ is not the index of the polarization but represents that it is a quantity for the antipodal correlations.}
    \begin{align}\label{eq:ca}
        C_A(t) 
        &\equiv \int_{-\infty}^\infty {\rm d}f \
        A^{\text{(D)}}_h(f) e^{-4\pi i f t} \nm \\
        &= -\int_{-\infty}^\infty {\rm d}f \
        S^{\text{(D)}}_h(f) e^{4\pi i f(\eta_0-t)} \,.
    \end{align}
Here, we have used Eq.~(\ref{eq:ah sh}) in the second line. 
The discussion seems to be parallel to the standard one. 
However, the problem is that $C_A(t)$ is extremely small compared to $C_S(\tau)$. 
Due to the phase factor $e^{4\pi i f\eta_0}$, 
the large contributions to $C_A(t)$ come from the modes with $f \lesssim 1/\eta_0 \sim 1/T_{\rm age}$. 
Moreover, when we take into account the $\hn$-dependent phase from the inhomogeneities, $e^{2\pi i f \eta_0 \hat{\phi}(\hn)}$, 
it will effectively work as the overlap reduction function: it suppresses contributions other than low-frequency modes with $f \lesssim 1/(\epsilon \eta_0)$. 
Here, $\epsilon$ represents the typical magnitude of $\hat{\phi}(\hn)$, i.e., the order of the scalar perturbations.
However, the signal $h_A(t,\hn)$ does not contain such low-frequency modes because any detector cannot detect GWs that do not vary over the observation time $T$. 
Therefore, 
$C_A(t)$ is extremely small and almost impossible to be detected. 
In Appendix \ref{sec:ca}, 
we explicitly give $C_A(t)$ for some examples of $S^{\text{(D)}}_h(f)$. 
We can also see that a filter for $t$, $Q_A(t)$, is not effective in increasing the sensitivity. 
Applying the filter $Q_A(t)$, we obtain
    \begin{align}
        &\int_0^T{\rm d}t \ C_A(t)Q_A(t) = \nm \\
        & \hspace*{1cm} -\int_{-\infty}^\infty {\rm d}f \
        S^{\text{(D)}}_h(f) e^{4\pi i f\eta_0}Q_{A;T}(-2f) \,,
    \end{align}
where $Q_{A;T}(f)$ is the short-time Fourier transform of $Q_A(t)$. 
For any choice of the filter function $Q_A(t)$, the support of $Q_{A;T}(f)$ has a width larger than $1/T \gg 1/\eta_0$. 
It is also impossible to cancel the phase factor $e^{4\pi i f\eta_0}$ by $Q_{A;T}(f)$.  
To achieve the cancellation, the inverse Fourier transform of $Q_{A;T}(f)$ should have a sharp peak at $t \simeq \eta_0 \sim T_{\rm age}$ 
but $t \sim T_{\rm age}$ is not included in the support of the inverse Fourier transform of $Q_{A;T}(f)$: $T_{\rm age} \notin [0,T]$. 
Therefore, the signal cannot be greatly enhanced for any filter function $Q_A(t)$. 

\subsection{\label{sec:ca} Some examples of $C_A(t)$} 

Here, 
we will compute $C_A(t)$,
    \begin{align}
        C_A(t) 
        &= -\int_{-\infty}^\infty {\rm d}f \
        S^{\text{(D)}}_h(f) e^{4\pi i f(\eta_0-t)} \nm \\
        &= -\int_{0}^\infty {\rm d}f \
        S_h(f) \cos[4\pi f(\eta_0-t)] \,,
    \end{align}
for some examples of $S_h(f)$. Here, we have rewritten the integral in terms of the single-sided spectral density (\ref{eq:single sh}). 

\subsubsection{Power-law spectrum} 
First, we consider the power-law spectrum $S_h(f) \propto f^{-\alpha}$ as usually assumed for the inflationary SGWB with $\alpha \simeq 3$. 
As we have commented in Sec.~\ref{sec:time}, the signal does not contain the low-frequency modes with $f \lesssim f_{\rm min} \equiv 1/T$ for the observation time $T$. 
Moreover, the spectrum should have an upper cut-off frequency $f_{\rm max}$ or the spectral index should satisfy $\alpha > 3$ in order that the total GW energy density is finite. Therefore, we consider the following integral,
    \begin{align}\label{eq:ca power}
        \int_{f_{\rm min}}^{f_{\rm max}} {\rm d}f \
        f^{-\alpha} e^{i f T_{\ast}} \,; \quad T_{\ast} \equiv 4\pi (\eta_0-t) \,,
    \end{align}
whose real part gives $C_A(t)$. 
Because the phase $fT_{\ast}$ is very large in the integral domain, 
we can estimate the integral by using the method of steepest descent. 
By deforming the path to $[f_{\rm min}, f_{\rm min} + i\infty] \cup [f_{\rm min}+i\infty, f_{\rm max} + i\infty] \cup [f_{\rm max} + i\infty, f_{\rm max}]$, 
we can evaluate the integral (\ref{eq:ca power}) as
    \begin{align}
        &\int_{f_{\rm min}}^{f_{\rm max}} {\rm d}f \ f^{-\alpha} e^{i f T_{\ast}} 
        = \nm \\
        &\qquad \int_{0}^{\infty} dp \ (f_{\rm min} + ip)^{-\alpha}e^{if_{\rm min} T_{\ast}  -T_{\ast} p} \nm \\
        &\qquad \quad - \int_{0}^{\infty} dp \ (f_{\rm max} + ip)^{-\alpha}e^{if_{\rm max} T_{\ast}  -T_{\ast} p} \,,
    \end{align}
where we have dropped the contribution from the path $[f_{\rm min}+i\infty, f_{\rm max} + i\infty]$ because $\lim_{p \to \infty} (f+ip)^{-\alpha}e^{-T_{\ast} p} = 0$. 
Since the damping factor $e^{-T_{\ast} p}$ suppresses contributions other than $p < 1/T_{\ast} \ll f_{\rm min} < f_{\rm max}$, 
we obtain
    \begin{align}
        &\int_{f_{\rm min}}^{f_{\rm max}} {\rm d}f \ f^{-\alpha} e^{i f T_{\ast}} 
        \simeq \nm \\
        &\hspace*{2cm} \frac{ f_{\rm min}^{-\alpha}e^{if_{\rm min}T_{\ast}}  - f_{\rm max}^{-\alpha}e^{if_{\rm max}T_{\ast}}}{T_{\ast}} \,.
    \end{align}
Therefore, $C_A(t)$ is estimated to be
    \begin{align}\label{eq:ca power result}
        &C_A(t) = \nm \\
        &\quad -\frac{S_h(f_{\rm min})\sin(f_{\rm min}T_{\ast}) - S_h(f_{\rm max})\sin(f_{\rm max}T_{\ast})}{T_\ast} \,,
    \end{align}
with $T_{\ast} \equiv 4\pi (\eta_0-t)$.
On the other hand, $C_S(0)$ given in Eq.~(\ref{eq:cs0}) is estimated to be
    \begin{align}\label{eq:cs power result}
        C_S(0) = \frac{S_h(f_{\rm max})f_{\rm max} - S_h(f_{\rm min})f_{\rm min}}{1-\alpha} \,.
    \end{align}
Comparing Eq.~(\ref{eq:ca power result}) with Eq.~(\ref{eq:cs power result}), we find that the antipodal contribution is suppressed at least by the small factor $1/(f_{\rm min}T_{\ast}) \sim T/T_{\rm age} = {\cal O}(10^{-10})$ compared to the standard one. 

\subsubsection{Gaussian spectrum} 
Next, we consider the Gaussian spectrum $S_h(f) \propto \exp[-(f-f_\ast)^2/2\sigma_f^2]$ as an example of a spectrum with a peak. We consider the following integral,
    \begin{align}\label{eq:ca gauss}
        \int_{-\infty}^{\infty} {\rm d}f \
        e^{-\frac{(f-f_\ast)^2}{2\sigma_f^2}} e^{i f T_{\ast}} \,; \quad T_{\ast} \equiv 4\pi (\eta_0-t) \,,
    \end{align}
where we have extended the integral domain by assuming that the peak is sufficiently sharp. 
This integral can be estimated as
    \begin{align}
        \sqrt{2\pi} \sigma_f e^{if_{\ast} T_{\ast}} e^{-\frac{\sigma_f^2 T_{\ast}^2}{4}} \,,
    \end{align}
and thus $C_A(t)$ is
    \begin{align}\label{eq:ca gauss result}
        C_A(t) = -\sqrt{2\pi} \sigma_f S_h(f_\ast) \cos(f_{\ast} T_{\ast}) e^{-\frac{\sigma_f^2 T_{\ast}^2}{4}} \,,
    \end{align}
with $T_{\ast} \equiv 4\pi (\eta_0-t)$. 
On the other hand, $C_S(0)$ given in Eq.~(\ref{eq:cs0}) is estimated to be
    \begin{align}\label{eq:cs gauss result}
        C_S(0) = \sqrt{2\pi} \sigma_f S_h(f_\ast) \,.
    \end{align}
Comparing Eqs.~(\ref{eq:ca gauss result}) and (\ref{eq:cs gauss result}), we find that the antipodal contribution is suppressed by the small factor $\exp(-\sigma_f^2T_{\ast}^2/4)$ compared to the standard one. 
Therefore, the antipodal contribution is extremely small unless the peak width $\sigma_f$ is much less than $1/T_\ast \sim 1/T_{\rm age}$.


\providecommand{\noopsort}[1]{}\providecommand{\singleletter}[1]{#1}%

\end{document}